\documentclass[a4paper]{elsart}
\usepackage{graphicx}

\begin{document}

\begin{frontmatter}

\title{The optical system of the H.E.S.S. imaging atmospheric
Cherenkov telescopes \\
Part II: mirror alignment and point spread function}

\author[2]{R.~Cornils}{,}
\author[1]{S.~Gillessen}{,}
\author[1]{I.~Jung}{,}
\author[1]{W.~Hofmann}{,}
\author[2]{M.~Beilicke}{,}
\author[1]{K.~Bernl\"ohr}{,}
\author[3]{O.~Carrol}{,}
\author[1]{S.~Elfahem}{,}
\author[2]{G.~Heinzelmann}{,}
\author[1]{G.~Hermann}{,}
\author[1]{D.~Horns}{,}
\author[1]{R.~Kankanyan}{,}
\author[1]{A.~Katona}{,}
\author[1]{H.~Krawczynski}{,}
\author[1]{M.~Panter}{,}
\author[4]{S.~Rayner}{,}
\author[1]{G.~Rowell}{,}
\author[2]{M.~Tluczykont}{,}
\author[2]{R.~van~Staa}

\address[2]{Universit\"at Hamburg, Institut f\"ur Experimentalphysik,
       Luruper Chaussee 149, D-22761 Hamburg, Germany}
\address[1]{Max-Planck-Institut f\"ur Kernphysik, P.O. Box 103980,
        D-69029 Heidelberg, Germany}
\address[3]{Dublin Institute for Advanced Studies, 
        5 Merrion Square, Dublin 2, Ireland}
\address[4]{Durham University, 
        The Observatory, Potters Bank, Durham, UK}

\begin{abstract}
Mirror facets of the H.E.S.S. imaging atmospheric Cherenkov
telescopes are aligned using stars imaged onto the closed lid of the PMT camera,
viewed by a CCD camera. The alignment procedure works reliably
and includes
the automatic analysis of CCD images and control of the
facet alignment actuators. On-axis,
80\% of the reflected light is contained in a circle of
less than 1~mrad diameter.
The spot widens with increasing angle to the telescope
axis. In accordance with simulations, the spot size
has roughly doubled at an
angle of $1.4^\circ$ from the axis. The expected variation of spot size with
elevation due to deformations of the support structure 
is visible,  but is completely non-critical
over the usual working range. Overall, the optical
quality of the telescope exceeds the specifications.
\end{abstract}

\end{frontmatter}



\section{Introduction}

H.E.S.S. (High Energy Stereoscopic System) \cite{hessreview}
 is a system of large
imaging Cherenkov telescopes currently under construction in the
Khomas Highland of Namibia {at 1800 m a.s.l., $23^\circ 16'$\,S,
$16^\circ 30'$\,E.} The first telescope
commenced observations in the summer of 2002. Three additional
telescopes are well advanced in their construction, and
should be completed in 2004.
Apart from a large mirror area --
over 100 m$^2$ per telescope -- the design of the system emphasizes the
stereoscopic observation of air showers over a large field of
view of $5^\circ$ diameter, as required both for the study of
extended TeV gamma-ray sources and for survey tasks. 

The purpose
of this paper and a companion paper is to describe and document the 
optical system of the H.E.S.S. telescopes and its performance.
{This optical system is used to image} the Cherenkov light
generated by air showers onto the photomultipliers (PMTs)
of the camera. The optics should provide high throughput and
an imaging quality matched to the resolution of the PMT camera
with its 960 pixels of $0.16^\circ$ size. The point 
spread function of the optics influences the image shape and hence
both the angular resolution of the telescope system 
and the gamma/hadron separation on the basis of image morphology.
It is therefore of prime importance that the imaging properties of the
telescopes are well understood, and that they are stable
in time and over a range of operating parameters.
The companion paper \cite{part1} -- in the following referred
to as Part I -- covers the layout and the 
components of the telescope optical systems,  including the 
segmented mirror with its support structure, the mirror facets -- 380
for each telescope -- and  the Winston cone light concentrators 
in front of the PMT camera. In this Part~II we concentrate on the technique
used to automatically align the facets using the motorized
facet actuators, and present the resulting point
spread function. 
We also illustrate how the data gained in the alignment process can be
used to study the stability of the dish structure.

\section{Mirror alignment}

\begin{figure}
\begin{center}
\includegraphics[width=9.0cm]{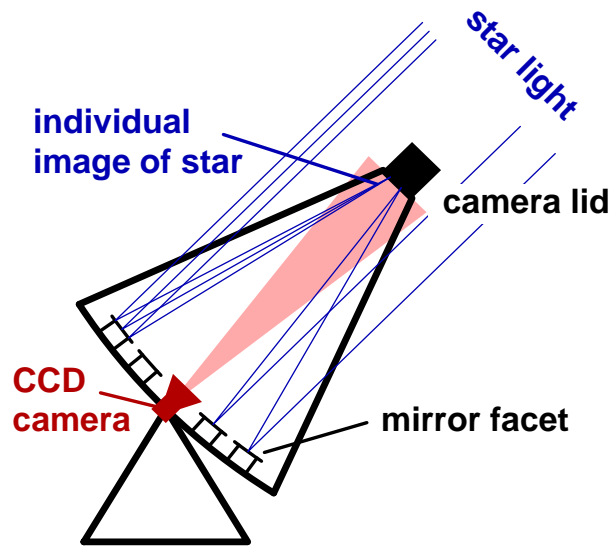}
\end{center}
\caption{Mirror alignment technique:  
the telescope is pointed towards an appropriate star whereupon all
mirror facets generate individual images of the alignment star in
the focal plane. Actuator movements change the location of the
corresponding light spot. This is observed by a CCD camera
viewing the lid of the PMT camera which acts as a screen.
}
\vspace{0.5cm}
\label{fig_method}
\end{figure}

\begin{figure}
\begin{center}
\includegraphics[width=12.0cm]{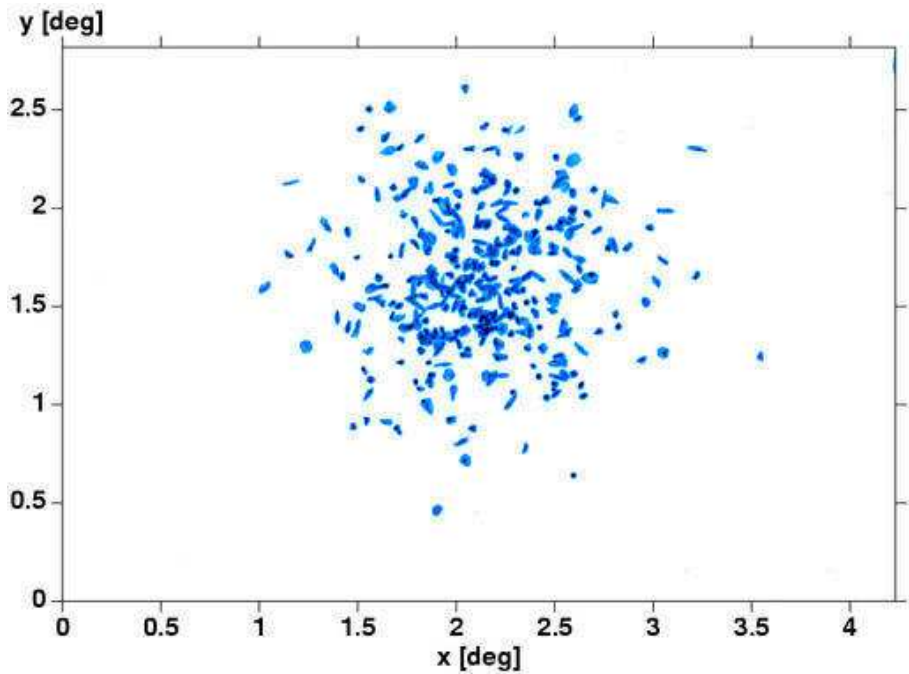}
\end{center}
\caption{Image of a star on the camera lid before alignment, using
a logarithmic intensity scale. 
The coordinates $x$, $y$ refer to the CCD image, translated into
degrees using the known focal length of the CCD optics. The origin is 
at the lower left corner of the CCD chip.
Each spot
corresponds to a mirror facet. The spread
of the spots reflects the precision
in the alignment of the facet support brackets on the dish, about
$0.5^\circ$, doubled due to reflection.}
\label{fig_unaligned}
\vspace{0.5cm}
\end{figure}

For optimum imaging quality the alignment of the 380 mirror facets
is crucial. The alignment uses the image of an appropriate star on the closed
lid of the PMT camera, 
viewed with a CCD camera at the center of the dish
(Fig.~\ref{fig_method}). 
{To serve as a screen, the 1.6\,m wide
camera lid is painted white and provides
on the optical axis a flat area of 60\,cm by 60\,cm free of bracings etc.}.
The individual facets are adjusted with
the goal of combining the star images of all facets into a single spot
at the center of the PMT camera.
The basic algorithm is as follows
\cite{align_icrc,align_memo}:   
a CCD image of the camera lid is taken. 
Before alignment, and with all facet actuators positioned
roughly at the center of their range, this image 
might appear as shown in Fig.~\ref{fig_unaligned}. A facet which is to   
be aligned is  
moved in both axes. The effects on the image are recorded
and provide all information required to move the corresponding
spot
to the center of
the main focus.
This procedure is repeated for   
all facets in sequence.
The alignment accuracy is constrained by the CCD resolution, the
actuator step size, and the mechanics of the facet supports, as well as by the design of the alignment algorithm.   

\subsection{The monitoring CCD cameras}

Given the large number of mirror facets on each H.E.S.S. telescope,
a reliable and automatic scheme for mirror alignment is required.
Such a scheme is implemented using the CCD camera to record the star images and to provide
feedback for the actuator movements.
The camera (``Lid CCD'') 
is located at
the center of the dish and is used to view images of stars on the
closed lid of the PMT camera. 
In addition, each H.E.S.S. telescope is equipped with a
a second CCD camera
(``Sky CCD'') which is mounted off-axis viewing the sky,
and which serves as a guide telescope. The main parameters of these cameras and
their optics are summarized in Table~1. The field of view
of the lid CCD should be adjusted such that the central region of the PMT
camera is covered and in addition at least three of eight positioning
LEDs mounted at the corners of the PMT array are viewed
\footnote{In the measurements with the first telescope - CT03 - only the central
region of the lid was covered; the CCD camera was readjusted later.}. These LEDs serve to 
monitor deformations of the camera masts. The two CCD cameras are optimized
differently. The lid CCD emphasizes a large dynamic range. During
the alignment process, images generated by individual facets as well
as the central spot with contributions from up to 380 facets must be resolved; the 
required frame rate is modest. The sky CCD, on the other hand, has a lower
intensity resolution, but high frame rate in order to
monitor telescope tracking continuously (during part of the measurements described here,
the fast Ap1E was not yet available, and a slower ST7 camera was
substituted). Both cameras are controlled using a PC running Linux,
and the images produced may be directed into the main data stream (during 
normal operation), and/or to a specialized mirror alignment process.
For the mirror alignment procedure only the lid CCD is used. 

\begin{table}
\begin{center}
\begin{tabular}{|l|c|c|}
\hline
& Lid CCD & Sky CCD \\
\hline
Optics & Nikkor lens & Vixen 120 NA S refractor \\
       & 180 mm, f/2.8 & 800 mm, f/6.7 \\
Camera type & Apogee Ap2Ep & Apogee Ap1E / Sbig ST7 \\
Pixels & 1536 x 1024 & 768 x 512 \\
Pixel size & 9.9'' & 2.3'' \\
Field of view & $4.23^\circ$ x $2.82^\circ$ & $0.51^\circ$ x $0.33^\circ$ \\
Full-frame readout & 45 s & 0.3 s (Ap1E) / 12 s (ST7) \\
ADC & 16 bit & 14 bit / 16 bit (ST7) \\
\hline
\end{tabular}
\caption{Characteristics of the two monitoring CCD cameras. During the 
initial work, an ST7 was used for the ``Sky CCD'', replaced later by
an Ap1E.}
\end{center}
\vspace{0.5cm}
\label{table_ccd}
\end{table}

\subsection{CCD image analysis}

For the mirror alignment stars between -1.5 and 2.0$^\mathrm{mag}$ were used,
with typical exposure times in the 0.2 to 16~s range;
the measurement of the point spread function was performed in the -1.5 to
4.1$^\mathrm{mag}$ range, with exposures from 0.7 to 100~s.
After alignment the image of a star typically contains a few $10^6$ electron
charges, spread over O(100) CCD pixels.
Signals are therefore very large compared to typical noise levels of 11
electrons (rms) per pixel.
During the alignment, isolated images generated by individual mirror facets
contain 2000 or more electrons.

{CCD image processing relies on the \emph{eclipse} library \footnote{The 
`ESO C Library for an Image
Processing Software Environment' provides services related to image handling,
filtering, dead pixel recognition, flatfielding, object detection, feature
extraction etc..} \cite{eclipse} from the European
Southern Observatory (ESO), supplemented by additional methods.
Details of the image processing differ between the mirror alignment procedure
-- where the emphasis is to define spot locations in a relatively short time --
and the measurement of the point spread function --
where the signal regions must be defined without discarding the tails of the
intensity distribution.

In the mirror alignment procedure, light spots generated by individual mirror
facets need to be identified.
The detection of these spots is based on the difference of two CCD images taken
in sequence for different positions of one of the alignment actuators of a certain
mirror facet.
The subtraction of the two images eliminates spots generated by other mirror facets, images of secondary stars, and
inhomogeneous background illumination (such as moon light partially shadowed by the
steel structure). Only the images generated by the mirror facet which
was moved remain, and are usually easy to identify.
Various consistency checks are applied during the analysis of the two CCD
images.
These include checks for intermediate tracking problems, changing weather
conditions, and accidental identification of image artefacts.
The difference method is very robust concerning image quality and therefore
the preparation of the raw images can be rather simple.
In a first step the (assumed constant) background level in the image
(i.e., electronics offset, dark current, background light)
is estimated by calculating the median of the pixel intensities,
and is subtracted.
The background subtracted images are cleaned by a 3x3 median filter in order
to eliminate single hot pixels. A 3x3 Gaussian filter smoothes
the images and simplifies the detection of image objects.

For the determination of the point spread function, the correct removal of the
background and the identification of pixels belonging to the light spots are of
prime concern.
Faint tails of the intensity distribution need to be included without the
introduction of a pedestal due to an underestimated background level.
In a first step, the spot location is estimated using the image object detection
algorithm of the eclipse library.
The background level is then determined by applying a Gaussian fit to the 
distribution of pixel intensities using a relatively small area around the
spot location.
Because of the good homogeneity of the CCD chip, a flat-fielding correction can
be omitted.
To determine the pixels belonging to the star image, a special spot extraction
algorithm has been developed.
The area surrounding the spot is divided into segments for which the total
intensities are calculated.
Segments with negative total intensities are discarded.
This is repeated for increasing segmentations of the remaining area.
After the isolation of the signal region the center of gravity of all associated
pixels is calculated.
In case the newly determined spot location differs by more than a quarter
of an image pixel from the previous location, the spot extraction is
reiterated.

Extensive tests were performed to ensure that the procedures to define signal
pixels do not introduce a bias in the determination of the point spread
function and image shape.
In addition, the results were verified using an independent analysis based on
an alternative segmentation algorithm \cite{jaehne,burt,gillessen_dipl}.}

\subsection{Location of focus}

For optimal imaging, a Cherenkov telescope should be focused at the height
of the shower maximum or somewhat higher \cite{ct_focus}. For a typical 
distance to the shower maximum of $D \approx 8$~km, this implies that the
focal plane should be positioned at 
$d \approx f/(1-f/D) \approx f + 28\mbox{ mm}$, 
where $f$ is the nominal
focal length. In case of the H.E.S.S.
telescopes, the mirror alignment was carried out using the images of stars
on the camera lid, positioned at $f = 15.00$~m. {The corresponding
focus for shower
images is then located about 30~mm behind the lid, and concident with
the entrance of the array of Winston cones mounted in front of the PMTs. 
The Winston cones then act as non-imaging light concentrators, funnelling
 light onto the photocathodes.
Therefore, after focusing star images on the camera
lid, the PMT camera is at the optimal focus for shower images.}

\subsection{Alignment algorithm}

The position of an individual light spot in the focal plane,
$\mathbf{x} \equiv (x_1,x_2)$, is given by a function 
\begin{equation}
\label{eq::alignExact}
\mathbf{x} = f(q_1, \ldots, q_6, \mathbf{a}) \, ,
\end{equation}
which depends on the position and orientation of the 
corresponding facet support, $q_i$, and on
the position of both actuators, $\mathbf{a} \equiv (a_1,a_2)$. 
Actuator positions are measured in units of counts of the
Hall sensors, which monitor turns of the alignment motors
(see Part~I for a detailed description of the facet support units, and of
the facet actuators and the
associated control electronics).
As the explicit expression for $f$ is rather complex and the exact values of
the $q_i$ are hard to determine, a linear approximation is used to relate 
$\mathbf{\Delta x}$ and $\mathbf{\Delta a}$:
\begin{equation}
\label{eq::alignApprox}
\mathbf{\Delta x} = \mathbf{T} \, \mathbf{\Delta a}
\, , \quad
\mathbf{\Delta a} = \mathbf{T^{-1}} \mathbf{\Delta x}
\end{equation}
The transformation matrix
\begin{equation}
\label{eq::alignTransform}
\mathbf{T} =
\left(
\begin{array}{cc}
\delta x_1 / \delta a_1	& \delta x_1 / \delta a_2	\\
\delta x_2 / \delta a_1	& \delta x_2 / \delta a_2
\end{array}
\right)
\end{equation}
is determined by driving both actuators individually for a fixed number of
 counts while imaging the change in $\mathbf{x}$ with the lid CCD camera. 
The components of $\mathbf{T}$ are correlated due to the geometry of the
facet support triangles, which are mounted on the dish with a fixed orientation:
\begin{equation}
\label{eq::alignRotate}
{\delta x_1 / \delta a_2 \choose \delta x_2 / \delta a_2} \simeq
\mathbf{R_{120}}
{\delta x_1 / \delta a_1 \choose \delta x_2 / \delta a_1}
\, , \quad
\mathbf{R_{120}} \equiv
\left(
\begin{array}{cc}
- 1 / 2		& - \sqrt{3} / 2	\\
\sqrt{3} / 2	& - 1 / 2
\end{array}
\right)
\end{equation}
Here, $\mathbf{R_{120}}$ accounts for the angle of $\sim120^\circ$ between the
two directions of movement.
The function $f$ is not strictly linear in $\mathbf{a}$ so the transformation
matrix itself is a function of $\mathbf{a}$ and its components vary by some
$\pm 10$\,\% for the whole range of the actuators.
The validity of a certain matrix $\mathbf{T}$ is thus confined to a small range.
To move a facet spot over a larger distance, an iterative procedure has to be used, with a compromise between 
precision and number of the iterations required (and hence the time
required for alignment). Various procedures (``alignment algorithms'')
have been studied. They differ, for example, in the effort made to determine
$\mathbf{T}$ or in the step sizes.

\begin{figure}[htb] 
\begin{center}
\includegraphics[width=8.0cm]{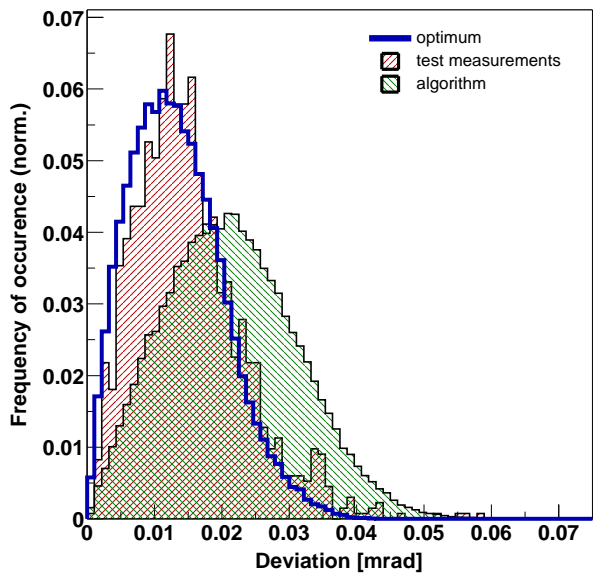}
\end{center}
\caption{
Alignment accuracy for individual light spots in the focal plane,
determined in a laboratory test setup. The narrow filled histogram shows
the experimental results, obtained by repeating the positioning cycle
many times.
The open histogram represents the theoretical optimum
based on simulations. It is in good agreement with the results
obtained from measurements with the test setup. 
The wider filled histogram represents the 
alignment accuracy of the algorithm chosen to align the telescopes,
which represents a compromise between speed and precision. The deviations
shown here refer to image coordinates, which are doubled due to 
reflection compared to the deviations in the orientation of
facets.
} 
\label{fig::alignDeviation} 
\vspace{0.5cm}
\end{figure}

To investigate the performance of different alignment algorithms, a simulation of the
 facet support mechanism and of the alignment process has been
developed.
{(With the intention of concentrating on the inaccuracies caused by the finite step
size of the actuators and by the uncertainty of the transformation matrix, the
influence of the CCD imaging with its finite pixel size and noise
was not simulated in detail.)}
For each algorithm investigated, $10^5$ mirror alignments have been simulated 
repeatedly for various distances of the mirror facet to the center of the dish. 
For reference, the alignment was first simulated using the exact
representation of $\mathbf{x}$ (Eq.~(\ref{eq::alignExact})) to determine the
maximum achievable accuracy. 
In this case,
the deviation from the nominal position is only due to the uncertainty of
one count in each actuator position and represents the limits of the hardware. 
The simulation resulted in a distribution for individual light spots in the 
focal plane with a rms deviation of $0.015$\,mrad and a worst case of 
$0.051$\,mrad (see Fig.~\ref{fig::alignDeviation}).
This is in good agreement with test measurements of the facet support
mechanics ($0.017$\,mrad rms, see Fig.~\ref{fig::alignDeviation}
and also Part~I).
After simulating the alignment accuracy for various 
algorithms, the algorithm described in the following section
 was chosen; its rms deviation of $0.023$\,mrad is only 
 marginally worse than the optimal case.

\subsection{Alignment procedure}

In addition to achieving a high alignment accuracy, the
algorithm should enable the fast and reliable realignment of the mirror
facets, in case the point spread function is observed to deteriorate
over time.
This is accomplished by splitting the procedure into two parts.
During the \emph{initial alignment}  certain parameters
are determined and stored in a database for further usage.
These parameters include the range and
position of the actuators and the final matrix to transform between CCD and actuator
coordinates, valid near the center of the camera.
The \emph{realignment} procedure does not redetermine these values
but simply retrieves them from the database.
{Since this transformation matrix depends primarily on the orientation
of the facets, it should not vary with time, and in fact no significant changes were
observed over several months.}
Following this approach, the time-consuming determination of these parameters has
to be performed only once. The time required to align mirrors is
to a significant extent determined by the CCD readout time. 
Another criterion was therefore
to minimize the number of CCD exposures and wherever possible, only preselected
CCD regions are read out.

The initial alignment procedure for a facet consists of 
several steps.
At first the correct connection to the corresponding node on the branch cable
is checked (see Part~I).
Next the range of both actuators is
determined (refered to as ``referencing'').
As the actuator motors are not equipped with encoders and only provide
information about the direction and amount of their travel, the lower stop is
used as an internal position reference.
The upstrokes of both actuators are determined by driving them to the lower
stop and measuring the range from there to the upper stop.
The valid range for positioning the actuators is 
then slightly reduced at both ends
(by 5\%) to avoid driving the actuators into the springs at the stops,
reducing the potential of mechanical failure.
Finally both actuators are driven to the lower end of their range to move the
facet image out of the field of view of the lid CCD camera.
Electrical problems and most types of mechanical defects will show up 
in these steps; the referencing is therefore also used to test the
modules.
These first two steps of the initial alignment procedure may take place during
daytime; no optical feedback is required.

The aim of the next step is to identify the individual image of a star (spot)
corresponding to the mirror facet to be aligned, and to determine a coarse
transformation matrix for positioning the spot near the nominal position (``coarse
alignment'').
The nominal position is defined by 
the center of gravity of the facets which have been already aligned
(the ``main spot'').
Due to the lack of a main spot for the first few facets, the spot of a laser
pointer is used instead.
The facets are aligned
in sequence, with only one unaligned spot in the CCD field of view at a given time;
images from other unaligned facets remain outside the field of view.
A good starting point for positioning the spot of a facet inside the 
field of view is to move
both actuators of a facet to the center of their range
(Fig.~\ref{fig_unaligned} shows the resulting spread for all facets).
Facets which have their image still outside of the CCD field of view
($\sim$2\%) must be pre-aligned with human intervention.
Next, a CCD image
is taken.
One of the actuators is then driven a predefined number of counts and
another CCD image is taken.
Here a 2$\,\times\,$2~CCD pixel binning is used which is sufficient for
deriving a coarse transformation matrix, while significantly reducing exposure and CCD
image readout times.
From the analysis of these images, two elements of the transformation
matrix are determined. Given these two elements, the other two can
be estimated with sufficient precision on the basis of Eq.~(\ref{eq::alignRotate}),
due to the fixed geometry of the facet supports.
Based on the measured spot location and the estimated matrix, both actuators
are then driven to move the facet spot to the location of the main spot.

\begin{figure}[htb]
\begin{center}
\includegraphics[width=10cm]{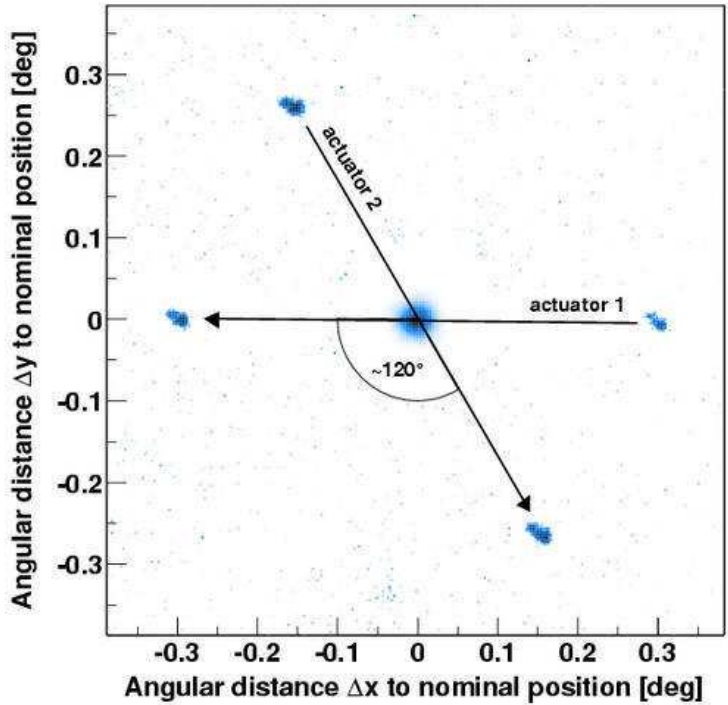}
\end{center}
\caption{
Superposition of four images taken by the lid CCD camera
during the fine alignment,
demonstrating the effect of the alignment actuators.
In turn, each of the two actuators of a facet
was moved a certain distance up and down, starting from a
position in the central spot (which is scaled down for better visibility of the
single spots). The resulting four displaced
spots serve to calibrate the actuator movements.
The angle of $\sim120^\circ$ between the 
two directions is due to the
geometry of the facet support triangle.
}
\label{fig::alignAlgo}
\vspace{0.5cm}
\end{figure}

The initial alignment ends with the determination of the transformation 
matrix with
maximum accuracy followed by the positioning of the facet spot at the center of
the main spot (``fine alignment'').
The procedure is similar to the coarse alignment; however,
the transformation matrix is derived for both actuators individually
and a 1$\,\times\,$1~CCD pixel binning is used.
To obtain an accurate transformation matrix for the area around the
main spot, the facet spot is placed at positions symmetrically
surrounding the main spot.
With this approach, the
resulting transformation matrix is essentially perfect for the 
vicinity of the main spot.
This step is illustrated in Fig.~\ref{fig::alignAlgo}.
The image readout time is reduced by retrieving only the relevant small 
sub-frame of the CCD image containing the spots.

The realignment procedure is essentially a fine alignment of a single actuator without
deriving a new transformation matrix. The spot of a given facet is moved
out of the main spot, located, and driven back to the exact center of the
main spot. {This technique requires two CCD exposures to locate the 
facet spot, before and after the facet is moved. An even faster
realignment was achieved by omitting the `before' exposure and using
the known transformation matrix to predict where to search for the spot
generated by the facet, after it is moved out of the main spot.
However, secondary stars accidentally overlapping the facet spot are then not
corrected and may result in slight alignment errors.}

{We note here that mirrors are aligned relative to the main spot
(or a laser spot for the first few mirrors). 
This procedure can be applied even if the telescope does not track the
star perfectly.
In any case, 
with a measured absolute pointing precision of about one arc-minute and
relative tracking deviations with respect to the average orbit of
a few arc-seconds, tracking of the H.E.S.S. telescopes is perfect for all practical
purposes. This holds in particular for the measurements of the 
point spread function discussed later; using bright stars, stable tracking over
a few seconds of exposure time is sufficient.}

Table~\ref{tab::alignSteps} summarizes the net duration for the different
steps of the alignment procedure.
\begin{table}[htb]
\begin{center}
\begin{tabular}{|l|l|l|}
\hline
step			& \multicolumn{2}{c|}{approximate net duration per}	\\
			& facet		& telescope	\\ \hline
connection		& 1 sec (daytime)	&		\\
referencing		& 8 min (daytime)	&
			\raisebox{1.5ex}[-1.5ex]{51 hrs}	\\
coarse alignment	& 90 -- 120 sec		& 		\\
fine alignment		& 90 -- 150 sec		&
			\raisebox{1.5ex}[-1.5ex]{24 hrs}	\\
total			& 12 min		& 75 hrs	\\
&&\\
realignment		& 45 -- 75 sec		& 6.5 hrs	\\
fast realignment	& 25 -- 40 sec		& 3.5 hrs	\\
\hline
\end{tabular}
\end{center}
\caption{Duration of the alignment procedure.
The first two steps can be performed during daytime while the two alignment
steps require darkness but allow for the moon to be partially visible.
The duration of the steps where optical feedback is required varies with the
brightness of the star used for alignment.
Under normal circumstances, the initial alignment needs about two weeks
and the realignment can be performed in one to two nights.}
\label{tab::alignSteps}
\vspace{0.5cm}
\end{table}
The steps requiring optical feedback can be performed during times where the
moon is partially visible, and do not interfere with the schedule for taking
very high energy gamma-ray data.

\section{Point spread function}

All mirror facets of the first H.E.S.S. telescope 
(for historical reasons labelled ``CT03'') were successfully aligned 
in January/February 2002; the second telescope (``CT02'') was aligned in 
November/December 2002. 
{In  construction, all four H.E.S.S. telescopes are  identical,
and should therefore exhibit identical mechanical and optical characteristics
\footnote{However, the weights required to balance the telescopes were
observed to differ
by O(100\,kg), most likely caused by tolerances in the wall thickness of
the camera arms.}}.
During the initial alignment of the telescopes,
the PMT cameras were not yet available; instead, dummies of
roughly the same weight and with front surfaces in the
same location as the real camera lid were installed.

\begin{figure}
\begin{center}
\mbox{
\includegraphics[width=8.0cm]{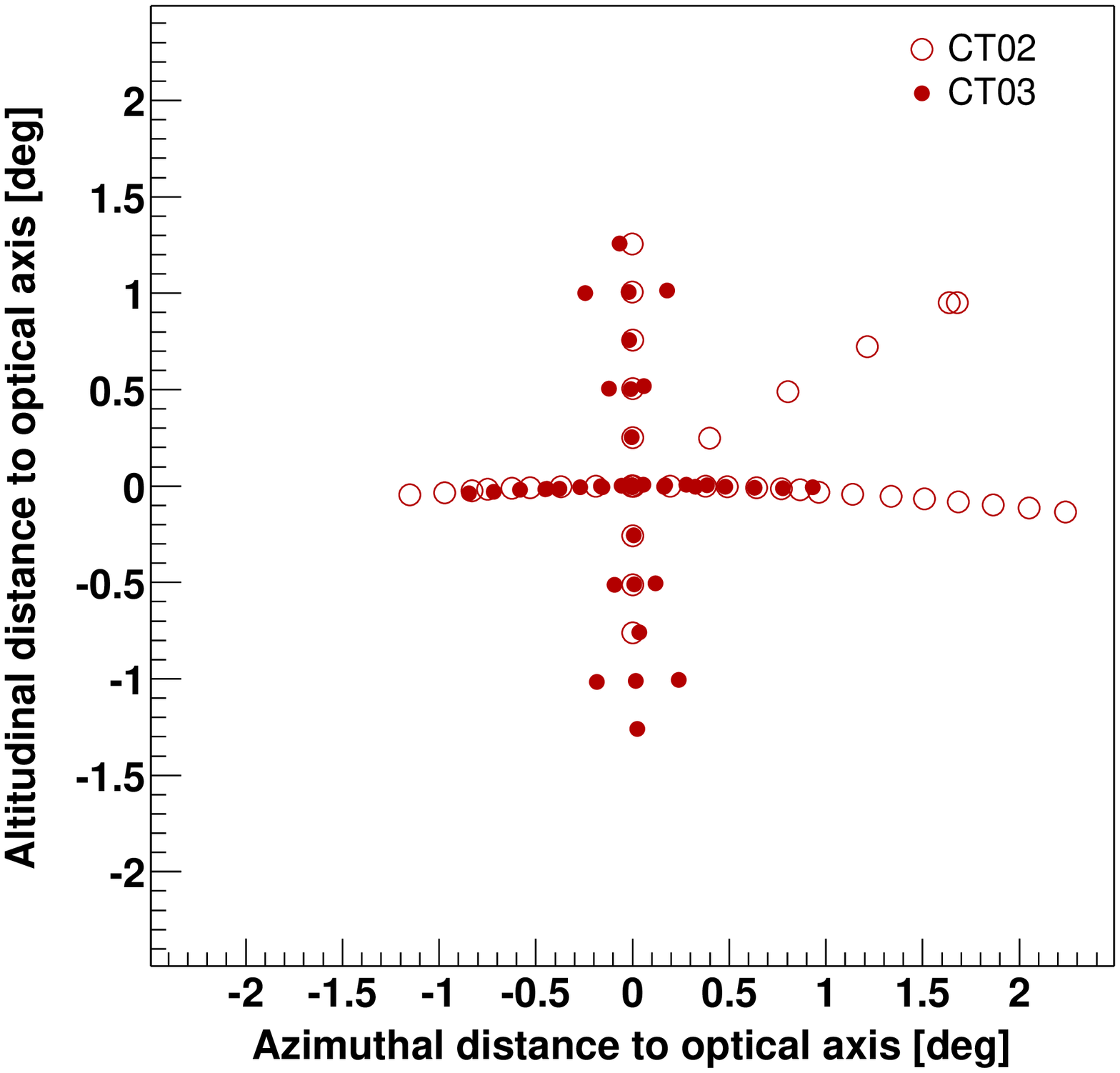}
\includegraphics[width=8.0cm]{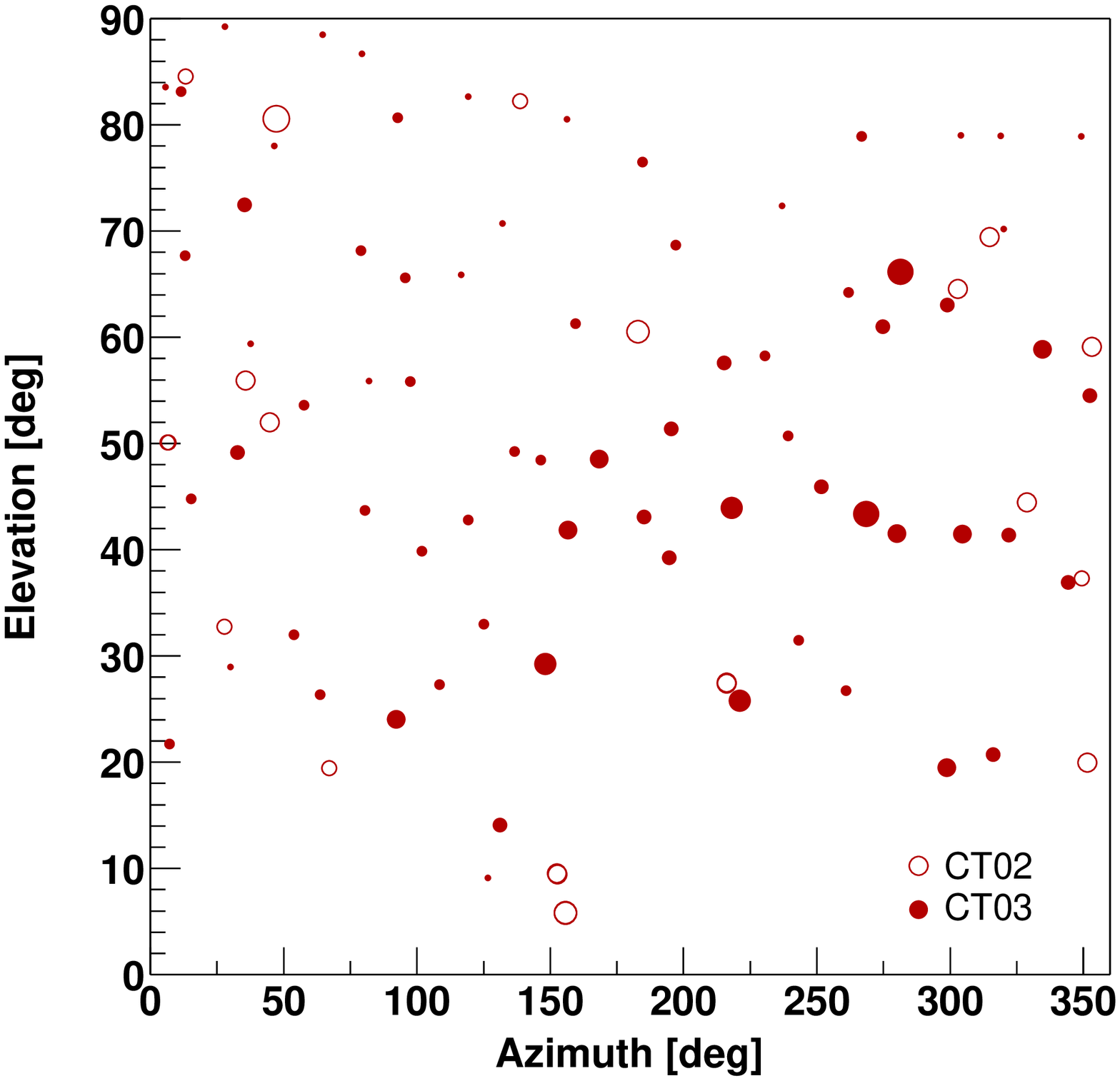}
}
\end{center}
\caption{Left: Locations in the field of view where the point
spread function was measured. The origin refers to the optical axis
of the telescope. Full symbols: first telescope (CT03), open symbols: 
second telescope (CT02).
Right: Elevation/azimuth values of telescope pointing where the
point spread function was measured. The azimuth coordinate increases from 
north to east. The size of the points indicates the brightness
of stars.}
\label{fig_alt_az}
\vspace{0.5cm}
\end{figure}

To study the point spread function after the alignment 
of the telescopes was completed, 
a variety of stars were used, covering a wide range 
of 
locations within the field of view (Fig.~\ref{fig_alt_az} left)
and of elevation and
azimuth angles for the telescope pointing (Fig.~\ref{fig_alt_az} right).
Because of the limited field of view of the lid CCD camera
of the first telescope, the
entire field of view of the PMT camera -- with $5^\circ$ diameter --
could not be covered in these measurements. For the second telescope,
the camera was pointed slightly off-axis, to cover both the
center of the field of view, and -- in one direction -- the edge.

\begin{figure}
\begin{center}
\includegraphics[width=10.0cm]{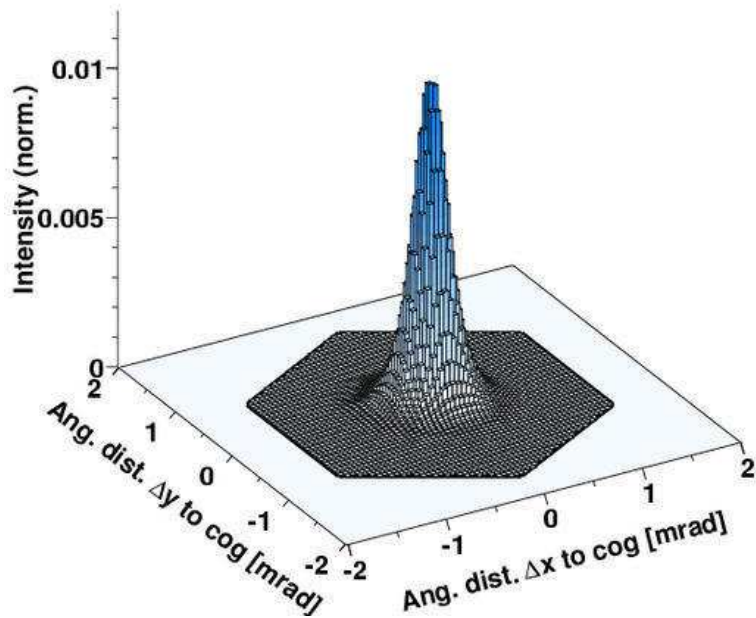}
\end{center}
\caption{Image of a star on the camera lid of CT03, as viewed by the lid CCD 
camera. The boxes correspond to CCD pixels, and the hexagonal border
indicates the size of a pixel of the PMT Cherenkov camera. {The image
was taken at an elevation of $70^\circ$, near the alignment range.}}
\label{fig_spot}
\vspace{0.5cm}
\end{figure}

The measured point spread functions of both telescopes are essentially
identical.
Fig.~\ref{fig_spot} shows a typical result, namely the CCD image of 
the image of a star on the camera lid, in relation to the size of a hexagonal
PMT pixel. The image spot is symmetrical, without pronounced substructure, 
and the width of the spot
is well below the PMT pixel size. 
As mentioned earlier, the signal charges in the CCD pixels are
very high, so that noise and statistical fluctuations in the 
signal are of no concern.

To quantify the width of the intensity distributions, 
different quantities are used.
{These include} the rms width $\sigma_{proj}$ 
of the projected (1-dimensional) distributions,
the rms width $\sigma$ of the two-dimensional distribution, and the radii
$r_{60}$ or $r_{80}$ of 
circles around the center of gravity of the image, containing 60\% or
80\% of the total intensity, respectively. For a symmetrical
Gaussian intensity distribution,
the following relations hold:

\begin{eqnarray}
\sigma & = & \sqrt{2} \sigma_{proj} \\
r_{60} & = & \sqrt{2 \ln{(5/2})}~\sigma_{proj} = \sqrt{\ln{(5/2)}}~\sigma \\
r_{80} & = & \sqrt{2 \ln{5}}~\sigma_{proj} = \sqrt{\ln{5}}~\sigma
\end{eqnarray}

On the optical axis, the point spread function is characterized by
values $\sigma = 0.33$\,mrad (CT02) and 0.32\,mrad (CT03)
(compared to an initial specification of 0.71\,mrad), 
$\sigma_{proj} = 0.23$\,mrad and 0.23\,mrad (compared to 0.50\,mrad), 
$r_{60} = 0.30$\,mrad and 0.28\,mrad (compared to 0.68\,mrad) and 
$r_{80} = 0.41$\,mrad and 0.40\,mrad (compared to 0.90\,mrad).

\subsection{Variation across the field of view}

Optical aberrations are significant in Cherenkov telescopes, due to
their single-mirror design without corrective elements
and their modest $f/d$ ratios. 
At some distance from the optical axis,
one therefore expects the width $\sigma$ of the point spread function to grow
linearly with the angle $\theta$ to the optical axis. On the axis,
the width of images is determined by the intrinsic optical
quality of the mirror facets, $\sigma_o$, by the aberrations
caused by the fact that most facets are inclined relative to the
optical axis, $\sigma_a$, and by the precision
$\Delta_{align}$ with which the individual mirror facets
are aligned. To a good approximation, the width of the point spread
function should hence be given by
$$
\sigma = (\sigma_o^2 + \sigma_a^2 + 4 \Delta_{align}^2 + c \theta^2)^{1/2}
$$
where the factor 4 accounts for the doubling of facet alignment errors
due to reflection. {The constant $c$ describes the aberrations
for off-axis rays and is proportional to the inverse square of 
the $f/d$ ratio of the telescope.}

\begin{figure}
\begin{center}
\includegraphics[width=14.0cm]{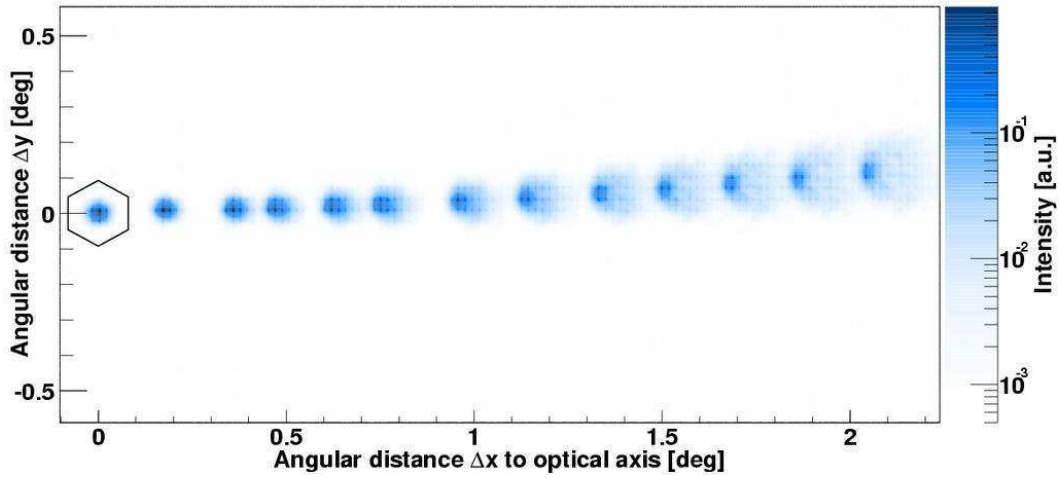}
\end{center}
\caption{Star images viewed on the camera lid of CT02, for different distances from
the optical axis. For comparison, the pixel size is indicated.}
\label{fig_off_axis}
\vspace{0.5cm}
\end{figure}

\begin{figure}
\begin{center}
\includegraphics[width=10.0cm]{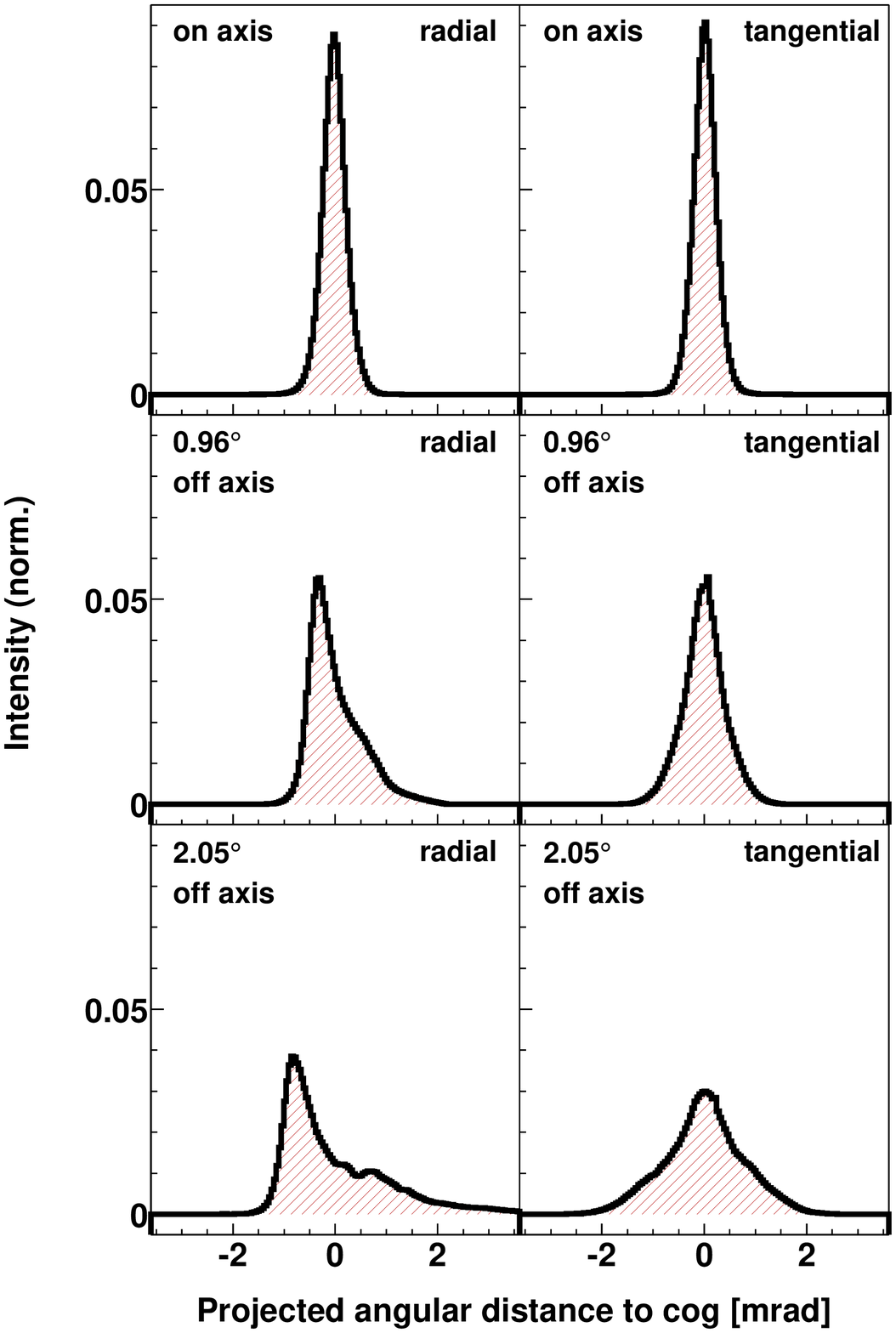}
\end{center}
\caption{Intensity distributions for images of stars on the camera lid
of CT02,
for stars on the optical axis (top row), $0.96^\circ$ off-axis (middle row)
and $2.05^\circ$ off-axis (bottom row), projected onto the radial
direction (left) and onto the tangential direction (right).}
\label{fig_proj}
\vspace{0.5cm}
\end{figure}

With increasing inclination to the optical axis 
 the spot is indeed observed to widen (Fig.~\ref{fig_off_axis});
it also develops 
an asymmetric tail away from the optical axis. A
more quantitative description of the spot shape is provided by 
projections of the 
intensity distribution on the radial and tangential directions, 
shown in Fig.~\ref{fig_proj}
for angles of $0^\circ$, $0.96^\circ$ and $2.05^\circ$ relative to the optical axis.

\begin{figure}
\begin{center}
\includegraphics[width=12.0cm]{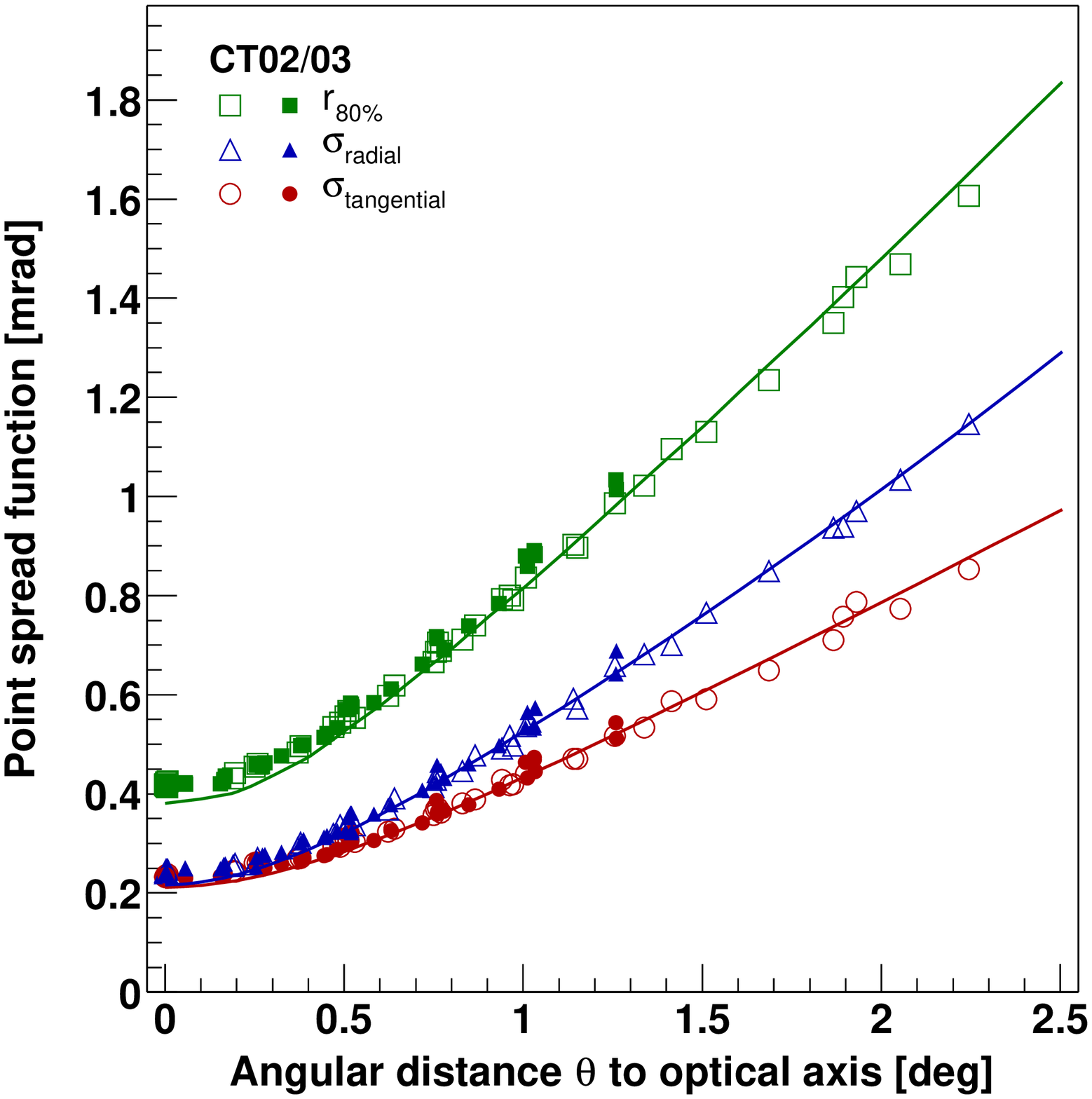}
\end{center}
\caption{Width of the point spread function as a function of the
angular distance $\theta$ to the optical axis, at elevations around $65^\circ$.
Different measures of the width of the point spread function are
shown: the rms width in the radial direction (triangles), the
rms width in the tangential direction (circles) and the radius
of a circle enclosing 80\% of the light, centered on the center
of gravity of the image (squares). 
Off-axis data were taken at different points of the field of view,
see Fig.~\ref{fig_alt_az}. 
Full symbols: first telescope (CT03), open symbols: 
second telescope (CT02).
{For CT03, the CCD camera used to view the spot was aligned to
provide a symmetric field of view with respect to the optical
axis, limiting its field of view to $1.5^\circ$. The CCD camera 
of CT02 has an asymmetric field of view and provides measurements
out to $2.3^\circ$.}
The lines indicate the result of simulations,
using the known {point spread functions of the mirror facets as input,
combined with the precision of the alignment algorithm as derived from
the simulations of the alignment process}.}
\label{fig_psf}
\vspace{0.5cm}
\end{figure}

For measurements near the elevation angle where the telescope
was aligned, around $65^\circ$, Fig.~\ref{fig_psf}
summarizes the spot parameters as a function of the 
angle $\theta$ to the optical axis.
In addition to $r_{80}$, the rms widths of the distributions
projected on the radial ($\sigma_{radial}$) and tangential
($\sigma_{tangential}$) directions are given.
The radial axis goes from camera center to the center of the spot, the
tangential axis is the corresponding orthogonal direction.
The measurements demonstrate that the spot width depends 
primarily on $\theta$; any other variation, depending on
the exact location in the field of view, is small and 
no other systematic trend is
found.
With increasing angle to the optical axis, the spot width increases,
following a shallow parabola at small angles. Later a transition
to a linear dependence is observed, as expected on the basis
of the modeling of the optical system (see Part I). The images
become slightly elliptical, elongated in the radial direction,
again as expected.
{The two telescopes show very similar point spread functions;
the last three points for $r_{80}$ of CT03 in Fig.~\ref{fig_psf}
are marginally 
above the corresponding values for CT02, but are still consistent
given the typical scatter of the measurements, and taking into account the
varying 
observing conditions, night sky brightness etc.}

To verify that the measured intensity distribution is quantitatively
understood, Monte Carlo simulations of the actual optical system
were performed, including the exact locations of all facets,
shadowing by camera masts, etc. 
As further input, the measured average spot size of the mirror facets
and the precision of the alignment algorithm were used. The latter is
govered primarily by the accuracy with which images of individual
facets can be located in the CCD images, which enters both directly
and indirectly via errors in the determination of the transformation
matrix and which results in an overall alignment uncertainty of 0.10
mrad for the individual spots. The errors in the location of images
were derived by moving some individual mirrors in steps along straight
lines and comparing expected and measured positions, and by comparing
the positions determined with different algorithms for image analysis.

The results are
included in Fig.~\ref{fig_psf} as solid lines, and are in
good agreement with the measurements. 
The simulations tend to slightly
underestimate the point spread function on axis. A likely explanation
is that the simulations use identical Gaussian point spread functions
for the individual facets, whereas some of the actual measured spots
are highly non-Gaussian, and in extreme cases even exhibit double
sub-spots.

\subsection{Variation with telescope pointing}

Given the weight of the facets and the size of the dish, it
would be non-trivial and quite costly to design a dish structure
which does not deform under the influence of gravity
when moved in elevation. The structure of the H.E.S.S. telescopes
represents a compromise between stiffness versus weight and cost.
Over the working range in elevation, about $30^\circ$ to
$90^\circ$, the influence of gravity-induced deformations should
be small compared to the intrinsic point spread function of
the mirror facets, specified by the requirement that 80\% of the
light is contained in a circle of 1~mrad diameter.

To test for variations of the spot size with telescope pointing,
the (most sensitive) on-axis point spread function was measured
over a wide range of pointings, see Fig.~\ref{fig_alt_az}.
No significant dependence of the point spread function on
telescope azimuth was detected at fixed elevation. This is non-trivial,
since in the H.E.S.S. telescopes additional stiffness is provided
by tensioning the dish between the two elevation towers. A deviation
from perfect flatness of the azimuthal rail could introduce an
azimuth-dependent modulation of the tension and hence of the shape
of the dish.

\begin{figure}
\begin{center}
\includegraphics[width=12.0cm]{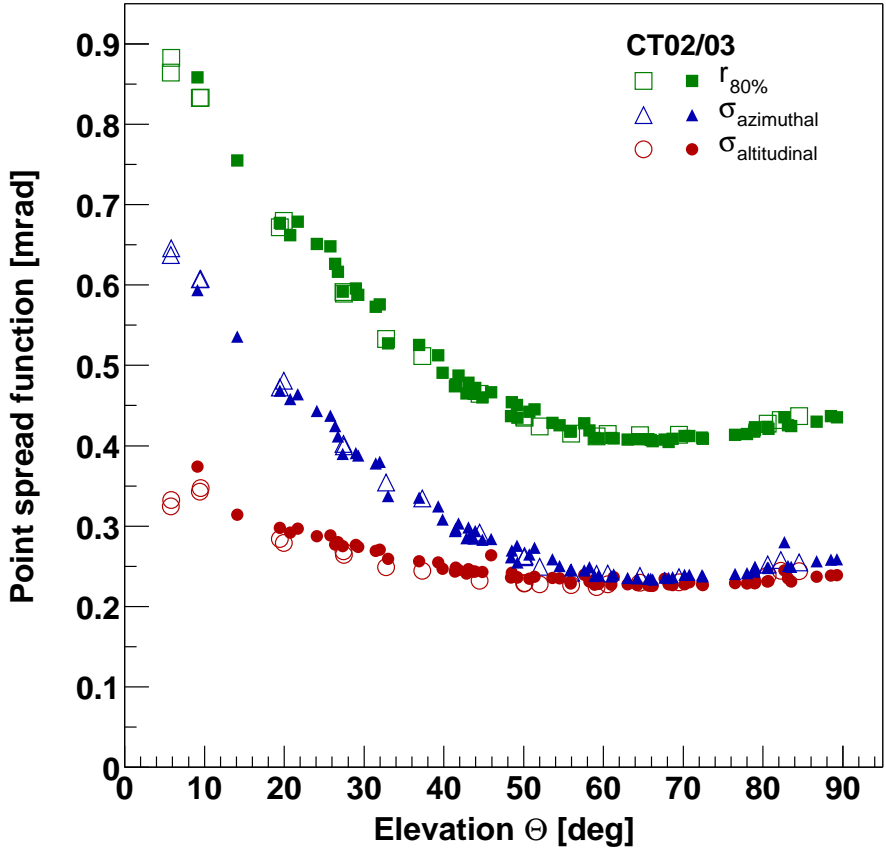}
\end{center}
\caption{Width of the point spread function as a function of 
elevation $\Theta$. Full symbols: first telescope (CT03), open symbols: 
second telescope (CT02).}
\label{fig_psf_alt}
\vspace{0.5cm}
\end{figure}

What is both predicted (see Part I) and observed is a 
variation of the point spread function with elevation, presumably
mainly due to deformations of the dish structure, but also with
small contributions from the facet support units. Fig.~\ref{fig_psf_alt}
illustrates how the measured spot width changes with elevation. The width 
is at a minimum around
$65^\circ$ to $70^\circ$, the angle at which the facets were aligned.
For the range of elevation angles most relevant for Cherenkov
observations (above $45^\circ$) the spot size $r_{80}$ varies by less than
10\%; at $30^\circ$ it is about 40\% larger than the minimum size.
At the alignment elevation the on-axis spot is essentially circular; 
at lower and higher
elevations the spot becomes elliptical and is narrower in the 
elevation direction as compared to the orthogonal direction,
along the telescope azimuth. {At low elevations, the
deviation between the spot widths along
its two main axes can reach 50\%.}

{To provide a description for use in simulations etc.}, the spot size
as a function of elevation $\Theta$
was parametrized by the function
$$
r_{80}(\Theta) = (r_{eff}^2 + d^2 (\sin\Theta-\sin\Theta_0)^2)^{1/2}
$$
with $r_{eff} = 0.41$\,mrad, $d = 0.96$\,mrad and $\Theta_0 = 66^\circ$.
The parameter $d$ refers to the deformation of
the dish relative to the alignment position; the measured value agrees within 25\% with the value
expected on the basis of finite element simulations of the
dish.
A more detailed discussion of the dish deformations will be given
in section 4.

\subsection{Long-term stability of the point spread function}

While a realignment of facets is possible within a relatively
short time, the telescope structure was designed for good 
long-term stability, to minimize drifts in the imaging 
characteristics. 

The point spread function of the first telescope (CT03) has been
measured several times over the course of nearly a year, first in February 2002
after the initial alignment, then in Summer 2002 after the  
PMT camera was mounted, and again
in December 2002 when the mirror of the second telescope (CT02) was aligned,
and in July 2003.
An increase in the width of the (on-axis) point spread functions of up to 15\% 
was observed between the first and the second measurement, 
in particular in $r_{80}$ and in the horizontal direction of the image.
This change is very likely caused by the
difference of about 100\,kg between the actual camera and the
dummy load used during the initial alignment; the forces 
generated by the camera arms represent a significant contribution
to the gravity-induced deformations of the dish. Since the
point spread function after installation of the camera was still
well below specifications, a realignment was not required.
Between the second, third and forth measurement, the point spread function
was stable.

\section{Study of deformations of the dish structure}

\begin{figure}
\begin{center}
\includegraphics[width=14.0cm]{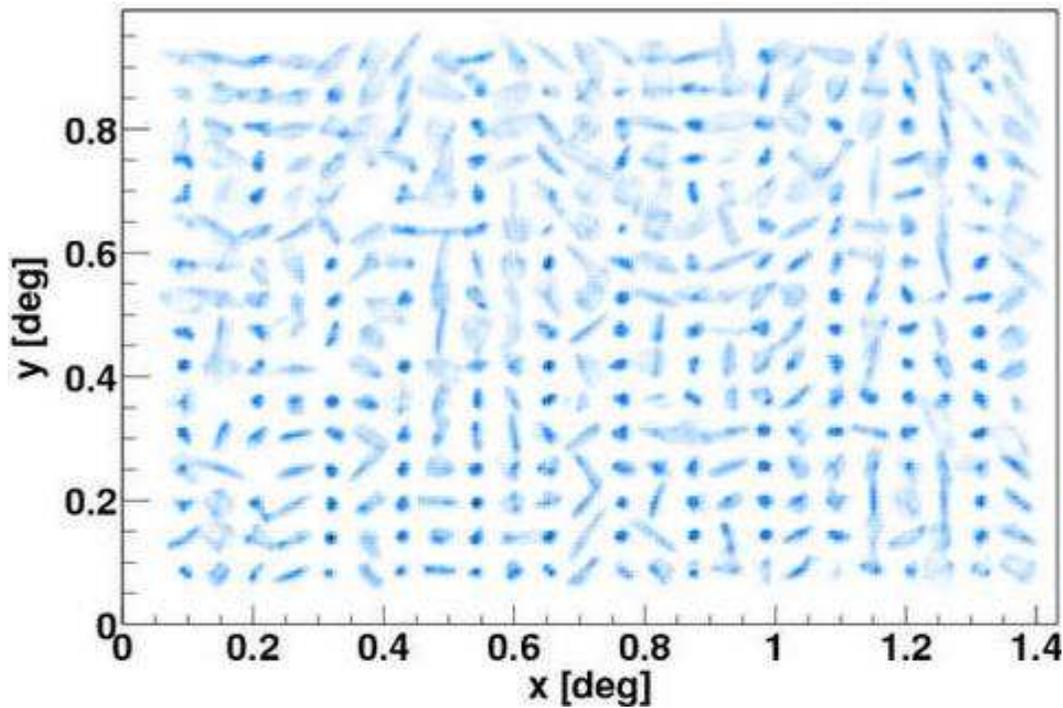}
\end{center}
\caption{Individual spots arranged in the form of a matrix. Each
spot corresponds to the image of the observed star generated by an 
individual
mirror facet; rather than combining all individual spots
at the center, the actuators were driven to generate this
specific pattern.}
\label{fig_matrix}
\vspace{0.5cm}
\end{figure}

The remote control of individual facets offers
additional options to study the mechanical characteristics of
the dish. Once facets are aligned -- i.e., the actuators are
referenced relative to an absolute system and the transformation
between actuator coordinates and image coordinates is determined
-- spots generated by individual facets
 can be arranged in arbitrary patterns. As an example,
Fig.~\ref{fig_matrix} shows the spots corresponding to
individual mirror facets arranged in the
form of a square matrix. Each element of the matrix is the
image of the observed star generated by one facet. One notices significant
differences between facets; some of the images are notably
elongated, while others are approximately circular spots. Part of these differences
are caused by variations in the quality of the mirror facets;
the bulk of the effect, however, are optical aberrations for
off-axis facets. Facets at the edge of the dish generate the
majority of the elongated images, and the orientations of the images
correlate with the position on the dish.

\begin{figure}
\begin{center}
\includegraphics[width=10.0cm]{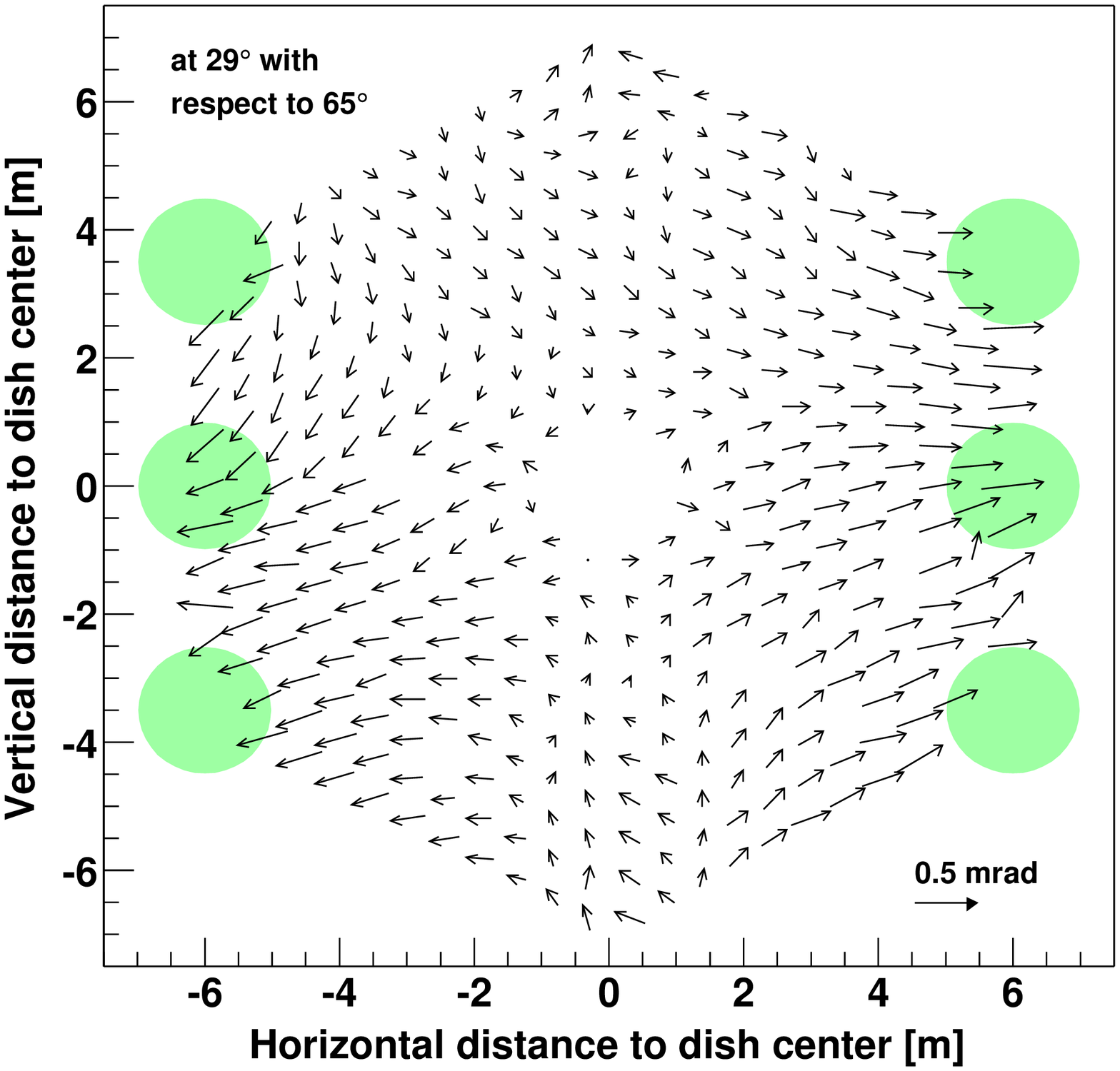}
\end{center}
\caption{Change in the orientation of facets
with elevation between 
$65^\circ$ and $29^\circ$
 elevation. The deflection
of a given facet is indicated as an arrow drawn at
the position of the facet on the dish; the length of the
arrow is proportional to the square root of the angular distance a facet
is deflected. Shaded areas indicate the regions where camera arms
are attached (top and bottom areas) and where the dish is supported
(middle).}
\label{fig_dishdef}
\vspace{0.5cm}
\end{figure}

The matrix pattern enables one to study deformations of the dish
structure in much greater detail, than is possible by observing only the
 effect on the overall point spread function. For this 
purpose, images of the matrix were taken for different elevations
of the dish. By tracking the relative movement of individual spots with
elevation, the deformation of the corresponding locations in the dish
can be determined. Fig.~\ref{fig_dishdef} shows
the deflection of facets between $65^\circ$ and $29^\circ$
elevation. 
Deformations are particularly
strong in the regions where the dish is supported and where the
camera arms are attached. 

\begin{figure}
\begin{center}
\includegraphics[width=10.0cm]{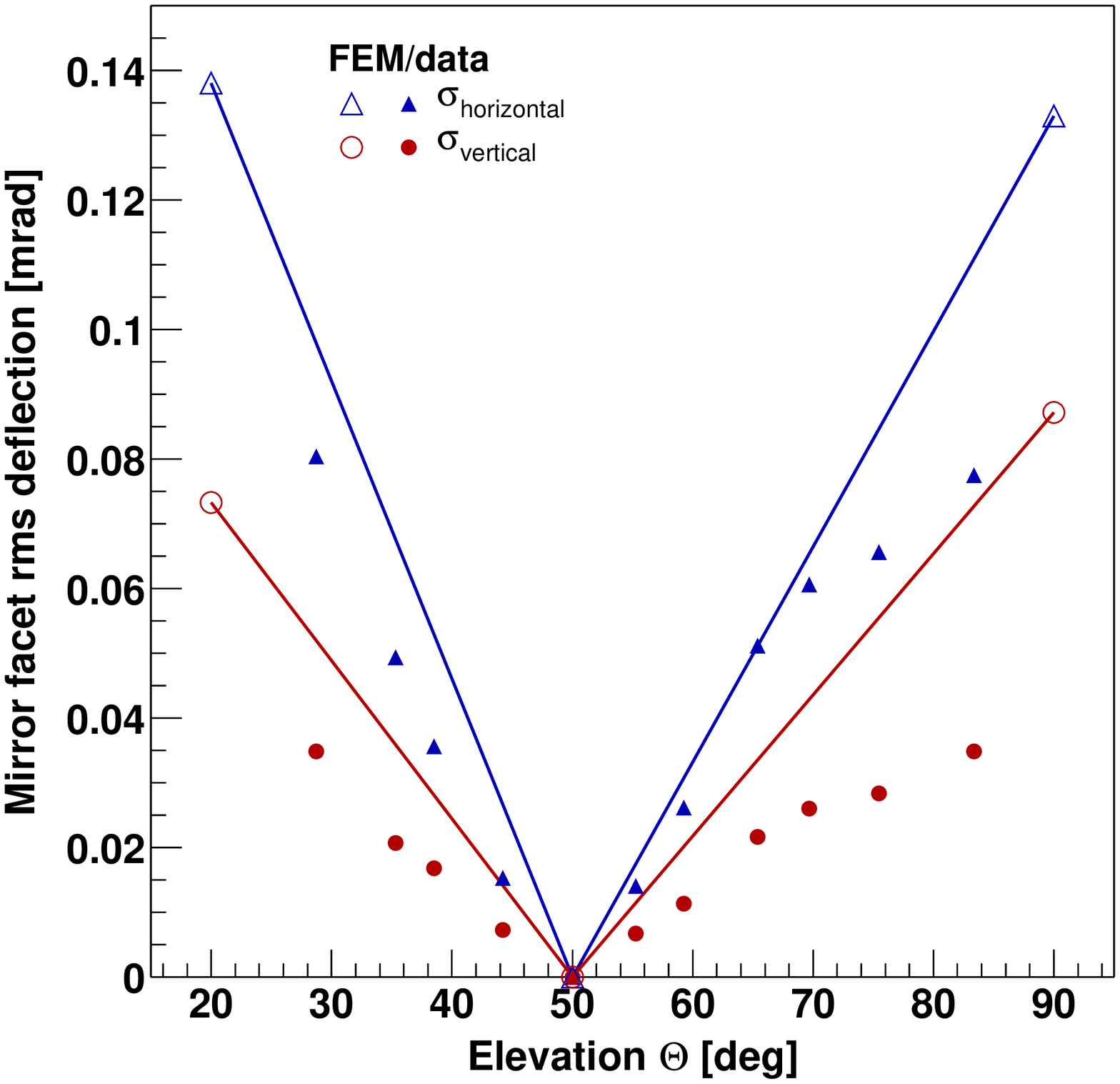}
\end{center}
\caption{Rms horizontal and vertical 
deflection of mirror facets relative to $50^\circ$
elevation, for data (closed symbols) and for the FEM model
of the dish (open symbols). Model calculations are only
available for $20^\circ$, $50^\circ$ and $90^\circ$; the
straight lines drawn between the model points should however
represent a reasonably reliable interpolation. The data were
obtained for CT03.}
\label{fig_mirrordef}
\vspace{0.5cm}
\end{figure}

The data on facet deflection allows a direct comparison between
the finite element (FEM) simulations and the actual performance of the 
dish.
In the design of the dish, FEM simulations were carried out for
elevations of $20^\circ$, $50^\circ$ and $90^\circ$
(see Part~I). In the simulations, the middle
elevation was defined as the reference point were the facets are
aligned \footnote{For the actual alignment, a higher reference
elevation around $65^\circ$ is used in order to optimize the
behaviour in the range between $45^\circ$ and $90^\circ$, where
most observations are carried out.}. To directly
compare data and FEM simulations,
facet deflections were determined relative to the facet pointing
at $50^\circ$ elevation. The rms deflection for all facets is
shown in Fig.~\ref{fig_mirrordef}, as a function of elevation,
separately for the horizontal and vertical deflections.

Data and simulation agree reasonably well; the dish actually
deforms less than predicted.
Given the differences between the actual dish and the
simplified model assumed in the simulations, such differences are to be expected
-- for some beams,
slightly modified wall thicknesses were used, weight was
added due to material in the nodes of the structure, walkways etc. 
Also, the FEM model made very conservative assumptions concerning the
stiffness of the nodes of the spaceframe backing the dish.

In summary, we conclude that the FEM simulations have proven
a reliable (or even conservative) tool for the design of the
H.E.S.S. instruments.

\section{Absolute pointing of the telescope}

The discussion was so far mainly concerned with the point spread
function, i.e. the width of the image. A second important quality
criterion of a Cherenkov telescope is the pointing precision.
{For strong gamma-ray sources, the location of the source on
the sky can be determined with a statistical precision of a few
arc-seconds. Systematic errors -- resulting, for example, from deviations
of the telescope pointing -- should be reduced to a similar level.
This is non-trivial, since image information obtained during 
normal operation -- such as stars in the field of view -- cannot
be used to monitor telescope pointing with sufficient precision,
given the coarse pixel size. Cherenkov telescopes therefore have
to rely primarily on shaft-encoder data for pointing (in the 
case of the H.E.S.S. telescopes, this information is
augmented by images from the optical guide
telescope -- the ``sky CCD''), and one needs to apply corrections
for effects such as the bending of the camera arms.} 

A detailed
discussion of this issue is beyond the scope of this paper
and will be addressed elsewhere. Here, we will only summarize the
main conclusions obtained at this time:
\begin{itemize}
\item Deformations of the camera arms, which result in a shift
of images relative to the PMT pixel matrix, are monitored by viewing
the set of LEDs at the edge of the PMT camera using the lid CCD;
the measured bending of the arms -- expressed in terms of the
image shift -- is well described by 
$b = 82.7" \cos{\Theta}$ .
\item Using a parameterization to relate measured shaft encoder
values to true pointing, a pointing precision of about 8"~rms is
obtained. The corrections account for the bending of the masts,
but also for nonlinearities of the encoders, alignment errors
of the telescopes azimuth and elevation axes, etc.
\item Using in addition the information from the sky CCD camera when
a star is in its field of view, pointing can be improved to
less than 3"~rms.
\end{itemize}

\section{Summary}

Using the first two telescopes of the H.E.S.S. system, an automatic alignment
technique for the mirror facets has been developed and tested, and
 the optical properties of
the telescopes were studied in detail. 

The alignment of the
380 facets proceeded to a large extent automatically. Operator
intervention was required only for a small fraction of facets where the actuator mechanics
had problems and needed to be exchanged, or where 
the initial spot was not within the field of view of the CCD camera. 
In total, about 75~h
are required in total to align all mirrors, about 2/3 of the
work being possible during daytime. A realignment 
is possible within one or two nights.

The point spread
function of a telescope depends on the optical design, on the quality of the 
individual mirror facets, on the precision of the alignment and on the
mechanical stability of the dish and the facet supports. 
{The
quality of the mirror facets and the precision of the alignment system
exceed specifications by a significant margin.
At the time of the design, 
 0.5~mrad was specified for the rms width of the one-dimensional projection 
of the spot, evaluated on the optical axis where the spot size is minimal (see Part I). This value should be compared 
to the measured width of 0.23~mrad. }

The point spread
function is expected to broaden with increasing distance from the
optical axis, and to vary slightly with elevation beause of
gravity-induced
deformations of the dish.
The point spread function, measured around $65^\circ$ elevation and 
characterized by the radius
$r_{80}$ of a circle containing 80\% of the light, and is well 
described by
$$
r_{80} = (0.42^2 + (0.71 \theta)^2)^{1/2}~~~\mbox{[mrad]}
$$
where $\theta$ is the angle to the optical axis in degrees. 
{The point spread functions are almost identical for the
two telescopes. Individual fits show a difference of 0.01 between
the telescopes for the first coefficient, and of 0.05 for the second.}
Monte
Carlo simulations using the measured facet characteristics as 
input predict
$$
r_{80} = (0.38^2 + (0.72 \theta)^2)^{1/2}~~~\mbox{[mrad]}
$$
indicating that the telescopes behave as expected.
With elevation $\Theta$, the on-axis spot size varies roughly as
$$
r_{80} = (0.41^2 + 0.96^2 (\sin\Theta - \sin66^\circ)^2)^{1/2}~~~\mbox{[mrad]}
$$
where $\Theta$ is given in degrees. This spot size
should be compared with the PMT pixel size of 2.8\,mrad.

\section*{Acknowledgements}

A large number of persons have, at the technical level, contributed to
the design, construction, and commissioning of the telescopes. 
We would like to express
our thanks to the technicians and engineers from the participating
institutes for their devoted engagement, both at home and in the 
field, initially frequently under adverse conditions. The team on site,
T.~Hanke, E.~Tjingaete and M.~Kandjii provided excellent support. 
S. Cranz, the owner of the
farm Goellschau, has given valuable technical assistance.
We gratefully acknowledge the contributions of U.~Dillmann, 
T.~Keck, W.~Schiel of SBP, of H.~Poller of SCE and of R.~Schmidt
and F.~van~Gruenen of NEC, responsible for the design and the
construction of the telescope structures. We thank J. Hinton for 
his comments on the manuscript.
Telescope construction
was supported by the German Ministry for Education and Research BMBF.

\end{document}